\documentclass[%
 reprint,
 amsmath,amssymb,
 aps,
]{revtex4-1}

\usepackage{graphicx}% Include figure files
\usepackage{dcolumn}% Align table columns on decimal point
\usepackage{bm}% bold math
\newcommand{\bea}{\begin{eqnarray}}
\newcommand{\eea}{\end{eqnarray}}
\newcommand{\beaa}{\begin{eqnarray*}}
\newcommand{\eeaa}{\end{eqnarray*}}

\newcommand{\del}[3]{\left#1 #3 \right#2}
\newcommand{\av}[1]{\del{<}{>}{#1}}

\begin{document}

\preprint{APS/123-QED}

\title{Comment: Causal entropic forces}% Force line breaks with \\

\author{Hilbert J. Kappen}
 \altaffiliation[Also at]{ Gatsby Computational Neuroscience Unit, UCL}%Lines break automatically or can be forced with \\
 \email{b.kappen@science.ru.nl}
\affiliation{%
Department of Neurophysics, %\textbackslash\textbackslash
Donders Institute for Brain, Cognition and Behaviour, \\
Radboud University
Nijmegen, The Netherlands
}%

\date{\today}% It is always \today, today,

\begin{abstract}
In this comment I argue that the causal entropy proposed in \cite{wissner-gross2013} is state-independent
and the entropic force is zero for state-independent noise in a
discrete time formulation and that the causal entropy description is incomplete in the continuous time case.
\end{abstract}

\maketitle
In a recent paper, \cite{wissner-gross2013} proposes a mechanism to explain the occurence of 
intelligent behavior. The proposal
is to consider a stochastic dynamical system and to compute the
entropy of trajectories over a finite time horizon, all starting
in the same initial state $x$. The dynamics is then a gradient flow
that maximizes this so-called causal entropy. 
%It is shown that the
%gradient is given by an expectation value of the first step of the
%trajectory weighted by the entropies of the remaining trajectories.
%The paper demonstrates the goal directed behavior of this dynamics
%on a number of interesting examples.
%

%\section{\label{sec:level1}First-level heading:\protect\\ The line
%break was forced \lowercase{via} \textbackslash\textbackslash}

%The proposal is intriguing and ambitious. It claims that one can generate intelligent behavior 'for
%free', in the sense that no specific objective has to be specified. 
%If true, this idea would more or less 'solve' the question of why
%intelligence exists. 

In this comment, I argue that the causal entropic force mechanism provides zero forces
for state-independent noise in any 
discrete time formulation with arbitrary small discretization $dt$ and that its description is incomplete in the continuous time case.

Consider a stochastic dynamical system of the form
\bea
dx_t = f(t,x_t) dt +  d\xi_t\qquad x_{t+dt}=x_t +dx_t \label{dynamics}
\eea
with $x$ an $n$-dimensional state vector, $f$ an arbitrary function and $\av{d\xi^2_t}=\nu(t,x_t) dt$ with $\nu(t,x)$ the noise covariance matrix.
By writing $x=(p,q)$, and allowing for the case that $\nu$ is not of maximal rank, this class of dynamical systems
contains all classical mechanical system with additive noise, in particular
the class of dynamical systems discussed in \cite{wissner-gross2013}.
We will discuss both the discrete time formulation with $dt$ a positive constant, in  which case we can set $dt=1$ without loss of generality. We also discuss the continuous time formulation with $dt\rightarrow 0$.

In the discrete time case,
consider 
a finite horizon time $T$ and consider trajectories $\tau=x_{1:T}$.
Let  $q(\tau|x_0)=\prod_{t=0}^{T-1} q_t(x_{t+1}|x_t)$
denote the probability to observe
a trajectory $\tau$ under the dynamics Eq.~\ref{dynamics} given an initial state $x_0$, with 
$q_t(x_{t+1}|x_t)$ a Gaussian distribution in $x_{t+1}$ with mean $x_t+f(t,x_t)$ and noise covariance matrix $\nu(t,x_t)$.
Define the Causal entropy in $x_0$ as
\bea
S(x_0)=-\int d\tau q(\tau|x_0)\log q(\tau|x_0)\label{entropy}
\eea
One can easily show that for any first order Markov process the path entropy is a sum
of contributions for individual times:
\bea
S(x_0)&=&s_0(x_0)+\sum_{t=1}^{T-1}
\int dx_{t} q_t(x_t|x_0)s_t(x_t)
\label{markov}\\
s_t(x_t)&=&-\int dx_{t+1} q_t(x_{t+1}|x_t)\log
q_t(x_{t+1}|x_t)\nonumber
\eea
with $q_t(x_t|x_0)$ the marginal probability to observe state $x_t$ at time $t$ given state
$x_0$ at time zero  and $s_t(x_t)$ is the entropy of the conditional distribution $q_t(x_{t+1}|x_t)$
\footnote{Note, that $q_t(x_t|x_0)=\int dx_{1:t-1} \prod_{s=0}^{t-1}q_s(x_{s+1}|x_s)$ is  non Gaussian for $t>1$ when $f(t,x)$ is a non-linear function of $x$.}.

Since $q_t(x_{t+1}|x_t)$ is Gaussian, $s_t(x_t)$
can be easily computed:
\bea
s_t(x_t)&=&\frac{1}{2}\log 2\pi \det \nu(t,x_t)+\frac{1}{2}
\eea
When $\nu(t,x_t)$ is not of maximal rank, the determinant is replaced by the so-called pseudo-determinant, 
defined as the product of the nonzero eigenvalues of $\nu(t,x_t)$.

%The path entropy becomes
%\bea
%S(x_0)&=&\frac{1}{2}\sum_{t=0}^{T-1}\int dx_t q_t(x_t|x) \log \det \nu(x_t) \label{gaussian}
%\eea
%where we have omitted an irrelevant constant term.

%
%
%which we can write as a sum of conditional entropies:
%\beaa
%S(x_0)&=&-\int dx_{dt:T} q(x_{dt:T}|x_0)\log q(x_{dt:T}|x_0)\\
%&=&-\int dx_{dt} q(x_{dt}|x_0)\log q(x_{dt}|x_0)-\\&&\sum_{t=dt}^{T-dt}
%\int dx_{t} q(x_t,t|x_0)\int dx_{t+dt} q(x_{t+dt}|x_t)\log
%q(x_{t+dt}|x_t)
%\eeaa
%where $q(x_t,t|x_0)=\sum_{x_{dt:t-dt}} q(x_{dt:t}|x_0)$ is the
%marginal probability to find the particle at time $t$ at position
%$x_t$, given that at time zero it was at $x_0$.
%Each of the entropy rate terms is given as
%\beaa
%&&-\int dx_{t+dt} q(x_{t+dt}|x_t)\log q(x_{t+dt}|x_t)\\&=&\int dx_{t+dt}
%q(x_{t+dt}|x_t) \left(\frac{1}{2}\log 2\pi \nu(x_t) dt
%+\right.\\&&\left. \frac{1}{2\nu(x_t) dt}(x_{t+dt}-x_t-f(x_t)dt)^2\right)\\
%&=&\frac{1}{2}\log 2\pi \nu(x_t) dt +\frac{1}{2}
%\eeaa
%Thus,
%\bea
%S(x_0)&=&\frac{1}{2}\log 2\pi \nu(x_0) dt
%+\frac{1}{2}\\&&+\sum_{t=dt}^{T-dt}\int dx_t q(x_t,t|x_0)
%\left(\frac{1}{2}\log 2\pi \nu(x_t) dt +\frac{1}{2}\right)
%\\
%&=&\frac{T}{2 dt}(\log 2\pi dt+1)+\\&&\frac{1}{2}\left(\log \nu(x_0) +
%\sum_{t=dt}^{T-dt}\int dx_t q(x_t,t|x_0) \log \nu(x_t)\right)
%\eea
%
When the noise is state independent, $\nu(t,x)=\nu(t)$, the causal entropy Eq.~\ref{markov} becomes $S(x_0)=\sum_{t=0}^{T-1}s_t$ because $\int dx_t q_t(x_t|x_0)=1$. {\em Thus, the causal entropy  is independent of
$x_0$ and the entropic force is zero.} This is true for arbitrary $dt>0$.
The examples that are reported in \cite{wissner-gross2013} are special case
of the dynamics Eq.~\ref{dynamics} with state-independent noise. 
Therefore, one cannot understand the reported intelligent behavior in these examples. 

%The path entropy does depend on the current state when the noise is
%state dependent. For instance, if the dynamics Eq.~\ref{dynamics} is ergodic and not explicitly time dependent, in the limit of $T\rightarrow \infty$ the causal entropy becomes
%\beaa
%\lim_{T\rightarrow \infty} S(x_0) = T \int dx q(x) s(x)
%\eeaa
%ic force is maximal for those $x$ for which
%$\nu(x)$ is maximal. The proposal to move the system to states of maximal causal entropy, thus becomes This seems a
%very disappointing conclusion.

Alternatively, one might consider a
 continuous time formulation. For arbitrary $dt$, 
 \beaa
 s_t(x_t)&=&-\int dx_{t+dt} q_t(x_{t+dt}|x_t)\log q_t(x_{t+dt}|x_t)\\
 &=&\frac{1}{2} \log 2\pi \det \nu(t,x_t)dt +\frac{1}{2}.
 \eeaa
In the limit $dt\rightarrow 0$, the path entropy Eq.~\ref{entropy} diverges and is not well-defined.
Instead, one may consider the {\em relative entropy}
\bea
K(x_0)&=&\int d\tau q(\tau|x)\log \frac{q(\tau|x)}{q_0(\tau|x)} 
\label{kl}
\eea
where 
$q(\tau|x_0)$ and $q_0(\tau|x_0)$ denote the distributions over trajectories under the dynamics Eq.~\ref{dynamics}
with drift terms $f(t,x)$ and $g(t,x)$, respectively and identical noise covariance $\nu(t,x)dt$.
 %Using $\ito$ calculus, denote $q(x_{t+dt}|x_t)$ and $p(x_{t+dt}|x_t)$ transition probabilities for the dynamics Eq.~\ref{dynamics}
%with drift terms $f_q(x)$ and $f_p(x)$, respectively and identical drift covariance $\nu(x)dt$. 
%which is well-defined when $q_0(x_{t+dt}|x_t)$ describes a stochastic dynamics of the form $dx_t=g(x_t)dt+d\xi_t$ and
%$\av{d\xi_t}=\nu(x_t)dt$ and $\nu(x)$ the noise same covariance matrix as Eq.~\ref{dynamics}.
One can show that
\beaa
K(x_0)=\frac{1}{2}\av{\int_0^T dt \ u(t,x_t)^T \nu(t,x_t)^{-1}u(t,x_t)}_q
\eeaa
with $u(t,x)=f(t,x)-g(t,x)$ and 
where $\av{}_q$ denotes expectation 
with respect to the distribution $q(\tau|x_0)$. $u(t,x)$ can be viewed as a control variable and $K(x_0)$ as the quadratic control cost 
\cite{kappen_prl05,kappen-gomez-opper-mlj2009}. $K(x_0)$ does depend on $x_0$ in this case and its gradient may provide the reported entropic force. The path (relative) entropy is minimised when $u(t,x)=0$. However, the interpretation of the causal entropy as a relative entropy depends on $g(t,x)$, which is not specified in \cite{wissner-gross2013}.

One further detail is the possible effect of walls or boundaries on the entropy production. When $\nu$ is state independent, $S(x_0)$ may still be state dependent when the walls are absorbing in which case probability is not conserved. In that case
the reported emergent behavior would be entirely the result of the interaction of the system with the walls.
However, in all examples in \cite{wissner-gross2013} it is explicitly stated that the collision with the walls are elastic. 
Such elastic collisions can be viewed as mirror images of the non-colliding trajectories and do not affect the entropy production $s(x_t)$.

In the case that 
not all degrees of freedom
are observable, the dynamics on the observed degrees of freedom is  no longer first order Markov. In that case, Eq.~\ref{markov} no longer holds and
the above conclusion may no longer be true. 
This may possibly explain the observed behavior in the
Tool Puzzle and Social Cooperation example, but not
the simpler examples Particle in a Box and Pole Balancing.
\cite{wissner-gross2013}, however, do not mention the necessity of partial observability for the results that they report.

\bibliography{/Users/bertkappen/doc/authors}

\begin{thebibliography}{4}%
\makeatletter
\providecommand \@ifxundefined [1]{%
 \@ifx{#1\undefined}
}%
\providecommand \@ifnum [1]{%
 \ifnum #1\expandafter \@firstoftwo
 \else \expandafter \@secondoftwo
 \fi
}%
\providecommand \@ifx [1]{%
 \ifx #1\expandafter \@firstoftwo
 \else \expandafter \@secondoftwo
 \fi
}%
\providecommand \natexlab [1]{#1}%
\providecommand \enquote  [1]{``#1''}%
\providecommand \bibnamefont  [1]{#1}%
\providecommand \bibfnamefont [1]{#1}%
\providecommand \citenamefont [1]{#1}%
\providecommand \href@noop [0]{\@secondoftwo}%
\providecommand \href [0]{\begingroup \@sanitize@url \@href}%
\providecommand \@href[1]{\@@startlink{#1}\@@href}%
\providecommand \@@href[1]{\endgroup#1\@@endlink}%
\providecommand \@sanitize@url [0]{\catcode `\\12\catcode `\$12\catcode
  `\&12\catcode `\#12\catcode `\^12\catcode `\_12\catcode `\%12\relax}%
\providecommand \@@startlink[1]{}%
\providecommand \@@endlink[0]{}%
\providecommand \url  [0]{\begingroup\@sanitize@url \@url }%
\providecommand \@url [1]{\endgroup\@href {#1}{\urlprefix }}%
\providecommand \urlprefix  [0]{URL }%
\providecommand \Eprint [0]{\href }%
\providecommand \doibase [0]{http://dx.doi.org/}%
\providecommand \selectlanguage [0]{\@gobble}%
\providecommand \bibinfo  [0]{\@secondoftwo}%
\providecommand \bibfield  [0]{\@secondoftwo}%
\providecommand \translation [1]{[#1]}%
\providecommand \BibitemOpen [0]{}%
\providecommand \bibitemStop [0]{}%
\providecommand \bibitemNoStop [0]{.\EOS\space}%
\providecommand \EOS [0]{\spacefactor3000\relax}%
\providecommand \BibitemShut  [1]{\csname bibitem#1\endcsname}%
\let\auto@bib@innerbib\@empty
%</preamble>
\bibitem [{\citenamefont {Wissner-Gross}\ and\ \citenamefont
  {Freer}(2013)}]{wissner-gross2013}%
  \BibitemOpen
  \bibfield  {author} {\bibinfo {author} {\bibfnamefont {A.}~\bibnamefont
  {Wissner-Gross}}\ and\ \bibinfo {author} {\bibfnamefont {C.}~\bibnamefont
  {Freer}},\ }\href@noop {} {\bibfield  {journal} {\bibinfo  {journal}
  {Physical review letters}\ }\textbf {\bibinfo {volume} {110}},\ \bibinfo
  {pages} {168702} (\bibinfo {year} {2013})}\BibitemShut {NoStop}%
\bibitem [{Note1()}]{Note1}%
  \BibitemOpen
  \bibinfo {note} {Note, that $q_t(x_t|x_0)=\DOTSI \intop \ilimits@ dx_{1:t-1}
  \DOTSB \prod@ \slimits@ _{s=0}^{t-1}q_s(x_{s+1}|x_s)$ is non Gaussian for
  $t>1$ when $f(t,x)$ is a non-linear function of $x$.}\BibitemShut {Stop}%
\bibitem [{\citenamefont {Kappen}(2005)}]{kappen_prl05}%
  \BibitemOpen
  \bibfield  {author} {\bibinfo {author} {\bibfnamefont {H.}~\bibnamefont
  {Kappen}},\ }\href@noop {} {\bibfield  {journal} {\bibinfo  {journal}
  {Physical Review Letters}\ }\textbf {\bibinfo {volume} {95}},\ \bibinfo
  {pages} {200201} (\bibinfo {year} {2005})}\BibitemShut {NoStop}%
\bibitem [{\citenamefont {Kappen}\ \emph {et~al.}(2012)\citenamefont {Kappen},
  \citenamefont {G{\'o}mez},\ and\ \citenamefont
  {Opper}}]{kappen-gomez-opper-mlj2009}%
  \BibitemOpen
  \bibfield  {author} {\bibinfo {author} {\bibfnamefont {H.~J.}\ \bibnamefont
  {Kappen}}, \bibinfo {author} {\bibfnamefont {V.}~\bibnamefont {G{\'o}mez}}, \
  and\ \bibinfo {author} {\bibfnamefont {M.}~\bibnamefont {Opper}},\
  }\href@noop {} {\bibfield  {journal} {\bibinfo  {journal} {Machine learning}\
  }\textbf {\bibinfo {volume} {87}},\ \bibinfo {pages} {159} (\bibinfo {year}
  {2012})}\BibitemShut {NoStop}%
\end{thebibliography}%
%\bibliographystyle{apalike} 


\end{document}